\documentclass[12pt]{article} 
\usepackage{amsmath}
\usepackage{amssymb}
 1                                        
\numberwithin{equation}{section}
\newcommand{\ing}{\not\! }
\newcommand{\1}{{\bf 1}\hspace{-3pt}\textrm{l}}
\newcommand{\ingl}{\not l}
\newcounter{li}
\newcommand{\li}{\stepcounter{li}{\theli.)}}
\begin{document}                                                              
\begin{center}                                                                
{ \large\bf All Electromagnetic Form Factors} 
\vskip 2cm 
{\ M. Nowakowski$^1$, E. A. Paschos$^2$ and J. M. Rodr\'{\i}guez $^1$,\\[0pt]
} $^1$Departamento de Fisica, Universidad de los Andes, Cra.1 No.18A-10,
Santafe de Bogota, Colombia\\[0pt]
$^2$Institut f\"ur Physik, Universit\"at Dortmund, D-44221 Dortmund, Germany

\vskip .5cm
\end{center}                             
\begin{abstract}
The electromagnetic form factors of spin-$\frac{1}{2}$ particles
are known, but 
due to historical reasons only half of them are found
in many textbooks. Given
the importance of the general result, its model independence, its
connection to discrete symmetries and their violations we made an effort
to derive and present the general result based only on the
knowledge of Dirac equation. We discuss the phenomenology connected directly with the form factors, 
and spin precession in external fields including time reversal violating terms.
We apply the formalism to spin-flip synchrotron 
radiation and suggest pedagogical projects.
\end{abstract}
\newpage

\section{Introduction}
In this article we present a detailed description of the electromagnetic form-factors with an
application to spin precession which includes time-reversal violating terms.
Our motivation to revisit an old subject \cite{gasiorowicz} 
stems from the fact that the field has become interesting again with new experimental results from 
Jefferson Laboratory. The topic is discussed in articles and textbooks which frequently do not include all 
terms, for instance the anapole form-factor. We include all possible form-factors and point out the various
assumptions introduced in their derivation. The presentation is conceptually simple, based on properties of
quantum mechanics. For the sake of simplicity we collect a list of algebraic relations in an appendix where we 
also demonstrate the derivation of Gordon-like identities.

There are also other reasons to justify this presentation. 
There are two types of spin-1/2 fermions: Dirac and Majorana \cite{majorana}. 
The former
carries beyond its mass and spin an internal global quantum number (like electric charge or
lepton number) which distinguishes particles from anti-particles.
The Majorana-particles are their own anti-particles.
The recent results on solar neutrino measurements, the atmospheric neutrinos and laboratory experiments
require neutrinos to have masses \cite{nureview}, which makes them good candidates as
Majorana fermions. 
It is interesting in this article to point out that the electromagnetic properties of Dirac and 
Majorana fermions differ greatly. The Dirac particles
couple to photon in four different ways (i.e. it has four form-factors), the Majorana particles
have only one coupling which in general is omitted in textbooks (see however, \cite{pal}) 
and if mentioned,
it is usually given without derivation.

Historically, the conservation of  
discrete symmetries like parity ($P$), time-reversal
($T$) and charge conjugation  or its combination $CP$ permitted two electromagnetic couplings.
These symmetries are known to be violated by the weak interaction and bring, in higher orders,
small violations of the discrete symmetries even to the electromagnetic couplings of fermions. 
In addition,
the special role that time reversal plays in the development of the matter-antimatter asymmetry of the
universe motivates a pedagogical review of the general case. Once upon a time the dictum was to preserve
discrete symmetries, but now that we know that they are violated at small, as well as, 
cosmological scales,
our attitude towards these symmetries has changed \cite{electdipole}.

Apart from the Dirac/Majorana nature of massive neutrinos and their
different electromagnetic properties, there remains the mystery of calculating any
of the form factors for extended particles. It was
long thought that the measurement of the first two electromagnetic
form-factors of the nucleons is a closed subject and 
only the issue of their parameterization remains to be discussed \cite{bosted}. 
Recent measurements at Jefferson Laboratory
\cite{discrepancy} have revealed a serious discrepancy: 
the old measurements do not agree with new ones using recoil polarization! 
The reason is still unknown, but it adds new excitement to the subject.

Last but not least, our motivation is to look for consistency and completeness in
teaching quantum mechanics. We start to learn or teach quantum mechanics by the 
assertion that the interpretation of the theory is only possible if given a 
wave-function $\phi({\bf x},t)$ whose the probability 
density $\rho=\phi^{\dagger}\phi$ is (i) positive 
definite (which in the form  $\phi^{\dagger}\phi$ is the case) and (ii) probability is conserved i.e.
there exist a current ${\bf j}$  such that
\begin{equation} \label{continuity}
\frac{\partial\rho}{\partial t}+{\bf \nabla} \cdot{\bf j} =0.
\end{equation}
 We can derive
${\bf j}$ given the Schr\"odinger equation for spin-0 particles. The electromagnetic
 current is then ${\bf J}=q{\bf j}$ with $q$ the charge of the particle: 
\begin{equation} \label{jem0} 
{\bf J}^{EM}_{spin-0}=\frac{iq}{2m}\left[\phi{\bf \nabla}\phi^{\dagger} -\phi^{\dagger}
{\bf \nabla}\phi -q{\bf A}\rho\right]
\end{equation}
with ${\bf A}$ the electromagnetic potential which appears in 
(\ref{jem0}) to maintain gauge
invariance. 
The limitation of the formalism is evident as we introduce additional `vectors' into the problem. For 
instance the introduction of spin allows a Pauli term
\begin{equation} \label{pauliterm}
{\bf \nabla}\times\left(\phi^{\dagger}
\mbox{\boldmath $\sigma$}\phi\right)
\end{equation}
to be added to the current because the divergence of a curl is always zero. The overall strength of 
this term in the Schr\"odinger formalism is arbitrary. It is determined by analyzing the contribution of the
new term to the energy of the system i.e. the Hamiltonian 
\cite{ll} or by a non-relativistic reduction of the Dirac current \cite{me1}.
The question now arises whether there are any additional terms to be added to the current.
In the non-relativistic theory it is difficult to devise the new terms, some of which are long and cumbersome.
It is more convenient to revert to the relativistic formulation where Lorentz covariance as a symmetry narrows
the number of possibilities. It is the aim of this article to present all such terms and then specify 
restrictions on them introduced by hermiticity, gauge invariance and the discrete symmetries of nature. In this 
approach we find form factors which are in principle known, but frequently omitted in books. One reason 
for this omission was, as mentioned above, the belief that discrete symmetries were exact. 
Because of the discovery that they are violated and the acceptance that time-reversal violation is 
a basic ingredient of the
big-bang in order to trigger baryogenesis, it is prudent to keep all terms and test their presence 
with experiment.

In section two we specify the general form of the electromagnetic current. We follow the steps mentioned above
by writing all possible terms and then use identities judiciously to eliminate many of them. Since the algebra
is cumbersome we discuss these identities in an appendix. We find the electromagnetic current can have four form 
factors, whose physical content in the non-relativistic limit is also discussed. In section three we show that
the spin precession in vacuum in external magnetic fields is due to the three  form factors at $q^2=0$. We derive
the spin precession relativistically \cite{bmt1, bmt2} and also keep the time-reversal violating contribution.
In section four we show how this contribution enters a Hamiltonian which can be used to calculate 
synchrotron radiation with a simultaneous change of the particle's spin. We close the article by summarizing the
results.   

Unless otherwise stated we use $\hbar=c=1$ throughout the paper.

\section{The Electromagnetic Current}
In order to arrive at the general expression for the current
we are going to consider the expectation value
\begin{equation}\label{expecvalue}
\langle p_1|j^{\mu}(x)|p_2 \rangle = e^{-i(p_2-p_1)x}\langle p_1|j^{\mu}(0)|p_2 \rangle
\end{equation}
where we have used a translation transformation
\begin{equation}
e^{iy\hat{P}}j^\mu (x) e^{-iy\hat{P}}= j^\mu (x+y)
\end{equation}
with $\hat{P}$ the four momentum operator \cite{bj2}. It is evident that we can write
\begin{equation}
\langle p_1|j^{\mu}(0)|p_2 \rangle = \bar{u}({\bf p}_1)\mathcal{O}^{\mu}(l,q)u({\bf p}_2)
\end{equation}
where $l^{\alpha} \equiv  p_1^{\alpha}+ p_2^{\alpha}$ and
$q^{\alpha} \equiv  p_1^{\alpha}- p_2^{\alpha}$
with $\mathcal{O}^{\mu}$ being an operator whose matrix element between the
spinors is a Lorentz vector. $\mathcal{O}^{\mu}$ is also a matrix acting on the spinors. We are interested
in writing the explicit form of $\mathcal{O}^{\mu}$. 

Let us first collect the requirements on our matrix element. Since $j^{\mu}$ is a 
Lorentz-vector $\mathcal{O}^{\mu}$ must also be a four-vector 
\footnote{When we mention four-vectors we mean that the matrix element
$\bar{\Psi}(p_1)\mathcal{O}^{\mu}\Psi(p_2)$ 
transforms like a four-vector.}
which we can ensure by working 
explicitly with tensors. This requirement is usually termed Lorentz-covariance (in our 
case it is manifest as we never handle any other quantities than tensors). The second condition 
is hermiticity i.e. $ j_{\mu}^{\dagger}=j_{\mu}$. This amounts to 
$\langle p_1|j^{\mu}(x)|p_2 \rangle=\langle p_2|j^{\mu}(x)|p_1 \rangle^*$ or 

\begin{equation}
\bar{u}({\bf p}_1)\mathcal{O}_{\mu}(l,q)u({\bf p}_2)= \bar{u}({\bf p}_1)\gamma^0\mathcal{O}_{\mu}^{\dagger}(l,-q)
\gamma^0u({\bf p}_2)
\end{equation}
which implies

\begin{equation}\label{hermiticity}
\mathcal{O}_{\mu}^{\dagger}(l,q)= \gamma^0\mathcal{O}_{\mu}(l,-q)\gamma^0
\end{equation}
Finally, current conservation (or gauge invariance) $\partial_{\mu}j^{\mu}=0$ can be recast into 
\begin{equation}\label{gaugeinvariance}
q^{\mu}\bar{u}({\bf p}_1)\mathcal{O}_{\mu}(l,q)u({\bf p}_2)=0.
\end{equation} 
Note that without the requirement of gauge invariance we can just derive what is called weak 
form-factor decomposition.

The next step in deriving the general relativistic electromagnetic current for 
spin-$\frac{1}{2}$ particles consists 
of collecting all possible four-vectors in terms of which we can parameterize $\mathcal{O}_{\mu}$. 
When we eliminate some 
candidates by using only Dirac algebra, especially the Gordon-like identities, we arrive at a current 
which is 
called weak current i.e. a non-conserved current (as we do not insist on gauge invariance) 
to which the massive 
vector bosons of weak interactions couple in analogy to electrodynamics, where $A_{\mu} j^{\mu}$ 
represents 
the coupling of the photon to the matter current. The fact that the weak current is not conserved has 
to do with the 
non-zero masses of the vector bosons mediating weak interaction \cite{note1}. Insisting in the next section  
on gauge invariance 
this gives us the electromagnetic current.

In order to construct the $4\times 4$ matrix $\mathcal{O}^{\mu}(l,q)$ we have 
at our disposal $\{ l^{\mu}, q^{\mu}\}$, 
the matrices in $\mathcal{S}$ (\ref{gammabasis}), the metric tensor $g^{\mu\nu}$ and the 
Levi-Cevita anti-symmetric 
tensor $\epsilon^{\mu\nu\alpha\beta}$. We define the first set by demanding that the Lorentz
index is carried by $q$ and $l$. Hence we get

\begin{equation}
\mathcal{O}_1= \left \{ q^\mu \1 , l^\mu \1 , q^\mu\gamma_5 , l^\mu \gamma_5\right \}.
\end{equation}
We could add to this a set 
\begin{equation}
\mathcal{O}_1'= \left \{ q^\mu \ing q  ,q^\mu \ingl , q^\mu \gamma_5 \ing q , q^\mu 
\gamma_5 \ingl , q^\mu \sigma^{\alpha\beta} q_{\alpha} l_\beta ,\textrm{ and } q^\mu 
\leftrightarrow l^\mu\right \}.
\end{equation}
but it is obvious that by using (repeatedly) (\ref{+eplanewave}) all terms in 
$\mathcal{O}_1'$ are proportional to the ones 
found in $\mathcal{O}_1$.

The next possible set of candidates is characterized by demanding that the Lorentz-index be 
carried by one of the 
matrices in $\mathcal{S}$ (\ref{gammabasis}). We have therefore
\begin{equation} \label{o2}
\mathcal{O}_2 = \left \{ \gamma^\mu ,\gamma_5 \gamma^\mu ,\sigma^{\mu\nu} q_{\nu},\sigma^{\mu\nu} 
l_\nu\right \}.
\end{equation}
Note that strictly speaking $\gamma_5 \sigma^{\mu \nu}q_{\nu}$ does not belong to the set (\ref{o2}) as
$\gamma_5 \sigma^{\mu \nu}$ is not linearly independent from the matrices in $\mathcal{S}$ (\ref{gammabasis}) due to
(\ref{propertiesofgamma}) (indeed this term is to be found in the next set below).
In the third set the Lorentz-index $\mu$ is carried by the Levi-Cevita tensor 
$ \epsilon^{\mu\nu\alpha\beta}$
\begin{equation} \label{o3}
\mathcal{O}_3= \left \{ \epsilon^{\mu\nu\alpha\beta} \sigma_{\alpha\beta}q_\nu,
\epsilon^{\mu\nu\alpha\beta}\sigma_{\alpha\beta}l_\nu,\epsilon^{\mu\nu\alpha\beta} 
\gamma_\nu(\1, \gamma_5) q_{\alpha} l_\beta, \epsilon^{\mu\nu\alpha\beta} q_\alpha l_\beta 
\sigma_{\nu\rho}q^\rho, \epsilon^{\mu\nu\alpha\beta} q_\alpha l_\beta \sigma_{\nu\rho}l^\rho \right \}
\end{equation}

The Gordon-like identities in (\ref{gordonlike2}) and (\ref{propertiesofgamma}) show that we can exclude
 $l^\mu \gamma_5$ (in favor of $\sigma^{\mu\nu}\gamma_5 q_\nu$ which is already included in 
$\mathcal{O}_3$,) 
$\sigma^{\mu\nu}l_\nu$ (in favor of $q^\mu$), and $l^\mu$ (in favor of $\gamma^\mu$ 
and $\sigma^{\mu \nu}q_{\nu}$). Furthermore the second identity in (\ref{gordonlike1})
and the rest of the  
Gordon-like 
identities involving the Levi-Cevita tensor in (\ref{gordonlike2})
demonstrate that only one candidate in 
$\mathcal{O}_3$ is independent.
Hence, taking everything together we arrive at six independent terms i.e. 

\begin{eqnarray}\label{sixfs}
\bar{u}({\bf p}_1)\mathcal{O}^{\mu}(l,q)u({\bf p}_2) &=& 
\bar{u}({\bf p}_1)\bigg\{f_1(q^2)q^\mu+f_2(q^2)q^\mu\gamma_5
+f_3(q^2)\gamma^\mu+f_4(q^2)\gamma^\mu\gamma_5 \notag\\
& & +f_5(q^2)\sigma^{\mu\nu}q_\nu+f_6(q^2)\epsilon^{\mu\nu\alpha\beta}\sigma_{\alpha\beta}
q_\nu\bigg \} u({\bf p}_2)
\end{eqnarray}

Indeed equation (\ref{sixfs}) represents the most general form-factor decomposition for the weak current if the
two fermions involved are on-shell and have equal masses. 
Note
that the form-factor, as indicated in (\ref{sixfs}) can depend only on a Lorentz-invariant quantity. 
Since 
$l\cdot q=0$ and $l^2+q^2=4m^2$ this quantity is $q^2$ (or alternatively $l^2$). 
Already the result (\ref{sixfs}) is 
widely used in particle and nuclear physics 
as it gives the general structure of the 
interaction (vertex) 
of a weak gauge boson with spin-$\frac{1}{2}$ matter.

The requirement of gauge invariance (\ref{gaugeinvariance}) is now easily
implemented on (\ref{sixfs}). 
It 
results into 
\begin{equation}
f_1 (q^2)q^2 +f_2 (q^2)q^2 \gamma_5+f_4(q^2)2m\gamma_5=0
\end{equation}
Since $\gamma^5$ and the unit matrix are linearly independent the above equation tells us that 
\begin{eqnarray}
f_1(q^2)&=& 0 \notag\\
f_4(q^2)&=& \frac{-f_2(q^2)q^2}{2m}
\end{eqnarray}
which leaves us with four electromagnetic form-factors. It is customary to express the final result 
through
 $F_i\quad(i=1,2,3,4)$ form-factors in the following form:
\begin{eqnarray}\label{fourff}
\bar{u}({\bf p}_1)\mathcal{O}^{\mu}(l,q)u({\bf p}_2) &=& \bar{u}({\bf p}_1)\left \{ F_1(q^2)\gamma^\mu
+\frac{i\sigma^{\mu\nu}}{2m}q_\nu F_2(q^2)+ i\epsilon^{\mu\nu\alpha\beta}
\frac{\sigma_{\alpha\beta}}{4m}q_\nu F_3(q^2)\right.\notag\\
& & \left.+ \frac{1}{2m}\left( q^\mu - \frac{q^2}{2m}\gamma^\mu \right)\gamma_5 F_4(q^2) \right\} u({\bf p}_2)
\end{eqnarray}
These agrees, for instance, with the results quoted in \cite{pal2} and \cite{kim} (given there without derivation).
Implementing the hermiticity condition (\ref{hermiticity}) gives us after using (\ref{daggergamma})
\begin{equation}\label{ffhermiticity}
F_i^*(q^2)=F_i(q^2)
\end{equation}
i.e. all form-factors in the parameterization chosen in (\ref{fourff}) are real (this is indeed the 
advantage of (\ref{fourff})). For instance,
$( i\sigma_{\alpha\beta})^{\dagger} = -i\sigma^\dagger_{\alpha\beta}=
-i\eta_0 [\sigma_{\alpha\beta}]\gamma^0\sigma_{\alpha\beta}\gamma^0 
= -i\gamma^0\sigma_{\alpha\beta}\gamma^0$. Equation (\ref{fourff}) is the most general relativistic current 
for the spin-$\frac{1}{2}$ 
fermion. It is worth discussing some of its properties.
\begin{list}{\li}{\setlength{\leftmargin}{0cm}%
\setlength{\labelsep}{6pt}\setlength{\itemindent}{18pt}}
\item In the derivation of (\ref{fourff}) we have considered the diagonal case i.e. the ket and bra 
in (\ref{expecvalue}) refer to the same particle with different momentum.
But, in principle, we could have started also with the off-diagonal case (say, electron and muon as the 
incoming and outgoing particles). In this case the result is
\begin{eqnarray}\label{offdiafourff}
\bar{u}_1({\bf p}_1)\mathcal{O}^{\mu}(l,q)u_2({\bf p}_2) &=& 
\bar{u}_1({\bf p}_1)\bigg\{(\gamma^\mu q^2-\ing q q^\mu )
\tilde{F}_1+i\sigma^{\mu\nu}q_\nu \tilde{F}_2\\
& & \left.+ i\epsilon^{\mu\nu\alpha\beta}\sigma_{\alpha\beta}q_\nu \tilde{F}_3+ \left( q^\mu - 
\frac{q^2}{m_1 + m_2}\gamma^\mu \right)\gamma_5 \tilde{F}_4 \right\} u_2({\bf p}_2) \notag
\end{eqnarray}
which satisfies all our requirements including 
$\tilde{F}_i=\tilde{F}_i(q^2)$ (Note that $l\cdot q=m_1^2-m_2^2$). Equations (\ref{fourff}) and 
(\ref{ffhermiticity}) 
represent the two results we have been looking for (see (3.1) and (3.3) in this context). 
The interaction of a spin-1/2 fermion with the 
four potential $A_\mu$ is $A_\mu(x)j^\mu (x)$ 
(which essentially is the energy density 
of the interaction) or in momentum space $\epsilon_\mu (\lambda ,q)\bar{u}({\bf p}_1)
\mathcal{O}^{\mu}u({\bf p}_2)$.

\item The decomposition (\ref{fourff}) and (\ref{offdiafourff}) is equally valid for point-like
particles like electron or muon and for extended particles like neutron and proton. It is valid 
for Dirac and Majorana fermions. We will however, see later that there is a difference between these 
two 
types of fermions as far as their electromagnetic properties are concerned. It is valid for charged and 
neutral fermions. The latter have a coupling to a photon either because of their extended nature or 
through their 
spin (spin-field interaction) as we will see later.

The difference between extended and point-like (elementary) particles is that extended 
spin-$\frac{1}{2}$ 
fermions have a priori the general form-factor structure given in (\ref{fourff}) and 
(\ref{offdiafourff})
as a result of their size. 
The functional form of $F_i(q^2)$'s is difficult to calculate from first principles,
because it depends on the internal structure of the proton and neutron.
For point-like 
objects we have a somewhat better undertaking. For point-like particles 
one starts with the coupling  
$eA^\mu \bar{\psi}\gamma^\mu\psi$ 
i.e. out of the four possibilities in (\ref{fourff}) we take only one. Higher order corrections in 
perturbation theory  
can produce then the structure 
(\ref{fourff}). The physical picture behind this is that the `bare' electron 
is always
accompanied with a cloud of virtual particles-antiparticles which makes it practically an extended
object.

\item What is the meaning of the form-factors? The easiest way to obtain some insight into an 
interpretation of 
$F_i$ is to couple the current to $A_\mu$ and take the non-relativistic limit. This is well known 
\cite{gasiorowicz,itzykson2,greiner}
and we quote only the results. One finds
\begin{equation}
F_1(0)=Q\,\,\,\,\, (\textrm{charge})
\end{equation}
leading to the interaction Hamiltonian
\begin{equation}
H^{NR}_{int}[F_1]=F_1(0)A_0
\end{equation}
where $A_0$ is the zeroth component of the four vector potential $A_{\mu}$. 
Similarly one can interpret
\begin{equation}
\frac{1}{2m}\left[F_1(0)+F_2(0)\right]= \mu\,\,\,\,\, (\textrm{magnetic moment})
\end{equation}
and deduce the Hamiltonian to have the form
\begin{equation} \label{int1}
H^{NR}_{int}[F_2]=-\mu\mbox{\boldmath $\sigma$}\cdot\bf{B}
\end{equation} 
This defines one of the most accurately measured quantities in physics. 
If we take $F_1(0)=e$ and $F_2$ proportional to $e$ we can define the
magnetic moment $\mu$  as $g(e/4m)$ where $g$ in non-relativistic quantum mechanics
is simply 2. If there is a deviation from this value, it is convenient to define a so-called
anomalous magnetic moment $a$ as
\begin{equation} \label{a}
a=\frac{g}{2} -1 =\frac{F_2}{e}
\end{equation}
For very recent experiments 
measuring the magnetic moment of the muon see \cite{muon}. 
The combined world average turns out to be
\begin{equation} \label{experiment1}
a^{\rm exp}_{\mu}= 11659203(8) \times 10^{-10}
\end{equation}
Whether this value agrees with theoretical predictions is still a matter of debate \cite{deRafael}.
The discussion of this paragraph refers to leptons, like electrons and muons.

The situation is more complicated for hadrons like proton and neutron. The first form-factors
at $q^2=0$ are again given by the electric charge of the particle, but the magnetic moments
cannot be predicted so easily and have anomalous values. In addition the $q^2$ dependence is determined
experimentally.
Both form-factors $F_1$ and $F_2$ have been measured over a wide range of
$q^2$. The results have been parametrized in different forms. For a quite recent discussion see
\cite{bosted}.
However, as already mentioned, there 
seems to be a discrepancy between these older results and recent experiments
which extract the two form-factors through polarization 
measurements \cite{discrepancy}. This is a very surprising outcome
as the physics of the two form-factors appeared to be a closed chapter, at least as far as 
their experimental determination is concerned. Certainly an explanation for this discrepancy is due.
One possible explanation for the differences is to include in the differential cross section
two photon exchange effects \cite{twophotons}.

The third form-factor is connected with the electric dipole moment via 
\begin{equation} \label{int2}
-\frac{1}{2m}F_3(0)= d\,\,\,\,\, (\textrm{electric dipole moment})
\end{equation}
with 
\begin{equation}
H^{NR}_{int}[F_3]=-d\mbox{\boldmath $\sigma$}\cdot\bf{E}
\end{equation} 
Up of now nobody has measured an non-zero $F_3$ at any value of $q^2$ or for any
particle.
This is not so surprising as we indeed expect this form-factor to be small. This expectation is based
on the fact that the electric dipole moment breaks the time reversal symmetry,
 as we will see below. We have indirect evidence 
from other experiments that such a violation occurs in nature, but the very same experiments indicate
that it has to be small. Nevertheless both theoretical physicists  \cite{theorydipole} and 
the experimentalists \cite{expdipole} think that it may be possible 
to find a non-zero $F_3$ in future.

Finally by the same methods one finds 
\begin{equation}
H^{NR}_{int}[F_4] \propto F_4(0)\mbox{\boldmath $\sigma$}
\cdot \left[{\bf \nabla}\times{\bf B}-\frac{\partial {\bf E}}{\partial t}\right]
\end{equation} 
and it is an excellent exercise to find the proportionality factor in the above equation. 
$F_4(0)$ is called (Zeldovich)
anapole moment and has a number of unusual properties. Firstly, note that by one of the Maxwell 
equations $H^{NR}_{int}[F_4]$ 
vanishes unless $\bf{j}\not = 0$ i.e. one of the sources of $\bf{E}$ and $\bf{B}$ is non-zero. 
This means that the 
coupling of the anapole moment to external electromagnetic fields $\bf{E}$ and $\bf{B}$ is of 
relevance only in matter!
Considered as the coupling of a photon to the fermion, the anapole coupling is zero if 
the photon is real 
(on-shell ; $q^2=0$ and $\epsilon_\mu q^\mu =0$). Hence, for instance, in bremsstrahlung processes 
$F_4$ does not contribute,
but in processes with off-shell photons it does; e.g. if the virtual photon is exchanged
between two fermions, the anapole moment will contribute to this process.
As is the case with $F_3$  we also lack a direct experimental evidence for $F_4$.
The reason is again to be searched in violation of one of the discrete symmetries. We can convince
ourselves that $F_4$ violates parity, a violation not as small as the one encountered in connection
with the time reversal. But in a real process like $ep \to ep$ where the form-factors are measured,
the parity violation through $F_4$ can interfere with a parity violation originating directly 
though weak interaction i.e. through a $Z^0$ exchange instead of a photon. 
The $Z^0$-proton-proton interaction
will be given by (\ref{sixfs}). It is hard to disentangle both contributions in a model
independent way.

\item What is the role of discrete symmetries in connection with $F_i$? We know from classical 
electromagnetism and 
non-relativistic quantum mechanics that under parity transformation $P$ and time-reversal transformation 
$T$ we have
\begin{align} \label{Ttrans}
\bf{E} & \xrightarrow{\ P\ }  -\bf{E}\quad  , & \bf{E} &\xrightarrow{\ T\ }  \bf{E} \notag\\
\bf{B} &\xrightarrow{\ P\ }   \bf{B}\quad , & \bf{B}&\xrightarrow{\ T\ }   -\bf{B}\\
\mbox{\boldmath $\sigma$}&\xrightarrow{\ P\ }\mbox{\boldmath $\sigma$}\quad,
&\mbox{\boldmath $\sigma$}&\xrightarrow{\ T\ }-\mbox{\boldmath $\sigma$}\notag
\end{align}
It is then obvious from our $H^{NR}_{int}[F_i]$ that the existence of non-zero $F_1$ and $F_2$ 
are compatible with $P$ 
and $T$ invariance. A non-zero $F_4$ signals clearly violation of parity  conservation and 
$F_3 \not = 0$ would tell 
us that time-reversal invariance is broken. Since the electromagnetic interaction by itself conserves 
both discrete 
symmetries we would expect $F_3=F_4=0$ (indeed, in most older textbooks only $F_1$ and 
$F_2$ are discussed). However, in 
reality the weak interaction which violates $P,\ T,$ and $C$ (charge conjugation) indirectly 
contributes to electromagnetic 
current via: photon$\rightarrow e^+e^-$ via electromagnetic interaction 
$\rightarrow e^+e^-$ interacting weakly
$\rightarrow$ on-shell $e^+e^-$. Hence the weak interaction contributes in an 
intermediate step.

\item Although we try to avoid the intricacies of Quantum Field Theory (QFT), one important 
issue is worth mentioning. In QFT 
the c-number field gets replaced by operators. If a scalar field representing a spin-0 particle has 
no global quantum number 
(including charge), i.e. it is anti-particle to itself, one expresses this fact by 
\begin{equation}
\mbox{\boldmath $\phi$}=\mbox{\boldmath $\phi$}^\dagger
\end{equation}
This, of course, has to do with available degrees of freedom to describe one or two states.
For fermions this condition is slightly more complicated as it reads
 \begin{equation}
\mbox{\boldmath $\psi$}=C(\bar{\mbox{\boldmath $\psi$}})^{\rm T}
\end{equation}
up to a global phase. Indeed denoting by $v({\bf p})$ the negative energy solution of the Dirac 
equation (\ref{dirac}) 
$C$ transforms $u$ from eq.(\ref{solve+e}) to $v$ via $v({\bf p})= C\bar{u}^{\rm T}({\bf p})$.
If $\mbox{\boldmath $\psi$}$ is a c-number, we get $(\bar{\mbox{\boldmath $\psi$}}
\Gamma\mbox{\boldmath $\psi$})=(\bar{\mbox{\boldmath $\psi$}}\Gamma\mbox{\boldmath $\psi$})^{\rm T}$.
When $\mbox{\boldmath $\psi$}$ is an operator we pick  up a minus sign in this process 
since 
fermionic operators anticommute . After a careful evaluation of 
$(\bar{\mbox{\boldmath $\psi$}}\Gamma\mbox{\boldmath $\psi$})^ \dagger$
using (\ref{daggergamma}) one gets
\begin{equation}
(\bar{\mbox{\boldmath $\psi$}}\Gamma\mbox{\boldmath $\psi$})=
-(\bar{\mbox{\boldmath $\psi$}}\Gamma\mbox{\boldmath $\psi$})^{\rm T}=
-\mbox{\boldmath $\psi$}^{\rm T}\Gamma^{\rm T}(\bar{\mbox{\boldmath $\psi$}})^{\rm T}=
-\eta_T[\Gamma]\bar{\mbox{\boldmath $\psi$}}
C^{\rm T}C\Gamma C^{-1}C^{-1}\mbox{\boldmath $\psi$}
=\eta_T[\Gamma]
\bar{\mbox{\boldmath $\psi$}}\Gamma\mbox{\boldmath $\psi$}
\end{equation}
where we used 
$C^{\rm T}=C^{\dagger}=C^{-1}=-C$. This equation
means that $(\bar{\mbox{\boldmath $\psi$}}\gamma_\mu\mbox{\boldmath $\psi$})=0$ and 
$(\bar{\mbox{\boldmath $\psi$}}\sigma_{\mu\nu}\mbox{\boldmath $\psi$})=0$ as $\eta_T[\sigma_{\mu \nu}]
=\eta_T[\gamma_{\mu}]=-1$.
Hence a Majorana fermion 
has only one 
electromagnetic moment: the 
anapole moment. This is in strong contrast to the Dirac case. 
Interestingly, this fact is again
connected to another symmetry of nature \cite{majoranaCPT}, that is the larger symmetry of CPT.
It is indeed a basic symmetry, since no violation of CPT has been reported
so far. 
\end{list} 
\section{Spin Precession}
The electromagnetic current for spin-$\frac{1}{2}$ fermions discussed in the last section 
has several important applications.
In the form $\int dx^3 A_{\mu}(x)j^{\mu}(x)$ it gives us the interaction energy and in quantum
field theory it results in all possible interaction terms of a photon with fermions, the so-called 
vertices. 
Yet there are other important applications of the form-factors found in $j^{\mu}$. One of them is
spin precession \cite{bmt1,bmt2} which, in principle, touches 
upon aspects of classical electrodynamics as we are
deriving this precession for the expectation value of the spin operator in a semi-classical limit.
The formulation of the form factors given in the previous sections allows us to discuss the spin precession
in its generality with `old' terms like the magnetic moment, but also with `new' terms like the electric
dipole moment which violates the time reversal invariance.

We have seen that the form-factors $F_{i=2,3,4}$ lead in a non-relativistic reduction to spin-field
interaction Hamiltonians (\ref{int1}) and (\ref{int2}). 
Restricting ourselves to $F_2$ and $F_3$ (taking also $F_4$ would force us
to consider the spin-precession in matter, a complication which we do not wish to consider  here) these
Hamiltonians also determine via the Heisenberg equation the time evolution of the operator
$\hat {\bf s} =(1/2) \mbox{\boldmath $\sigma$}$. Explicitly we obtain
\begin{equation} \label{Heisenberg}
\frac{d \hat{\bf s}}{dt}=2\mu \hat{\bf s} \times {\bf B}' + 2d 
\hat {\bf s} \times {\bf E}'
\end{equation}
where the primes indicate that the electric and magnetic fields values are taken in the rest frame of 
the particle. We set in this equation $\hbar=1$.

In electromagnetism we use (i) the Maxwell equations to determine the fields from
the sources and (ii) the Lorentz force which to determine the trajectory of the test charge. However, 
from the point of view of quantum mechanics the 
latter is  an expectation value of, say, velocity in the semi-classical approximation. Seen from this
perspective the equation for the expectation value of the spin in an external field has the same conceptual
status as the Lorentz force. We could add such a semi-classical equation for spin  as a third 
point (iii) 
to the other points above to encompass the
whole classical electromagnetism.

Denoting the expectation value of the spin by 
$\mbox{\boldmath $\xi$}$ we get from (\ref{Heisenberg})
\begin{equation} \label{Heisenbergclassical}
\left (\frac{d \mbox{\boldmath $\xi$}}{dt}\right)_{\rm rest frame}
=2\mu \mbox{\boldmath $\xi$} 
\times {\bf B}'
+2d \mbox{\boldmath $\xi$}\times {\bf E}'
\end{equation}
where in accordance with equation (\ref{Heisenberg}) the change of the expectation value
$\mbox{\boldmath $\xi$}$ with respect to time should be evaluated in the rest frame as indicated. 
The 
above equation is derived from a non-relativistic Hamiltonian and is therefore only a
non-relativistic form of a more general equation which we are looking for. Such relativistic generalization
calls also for the relativistic generalization of the concept of spin $s^{\mu}$.
In a similar way in which for a relativistic
concept of a four-vector force $f^{\mu}$ leads to $f^0={\bf f}\cdot{\bf v}$, one can also
show that $s^0={\bf s}\cdot {\bf v}$ or in other words
\begin{equation} \label{fourspin}
s^{\mu}p_{\mu}=s^{\mu}u_{\mu}=0
\end{equation}
where $u_{\mu}$ is the four-velocity. 
Equation (\ref{fourspin}) follows essentially from $s^{\mu}_{\rm rest frame}=(0,\mbox{\boldmath $\xi$})$.
With these provisions one can derive the relativistic
version of (\ref{Heisenbergclassical}) either from Dirac equation directly or by a similar method with which
we derived the relativistic current in previous sections.
This means that replacing $d \mbox{\boldmath $\xi$}/dt$ by $d s^{\mu}/d\tau$ 
with $\tau$ the proper time, we look for possible expressions for the right-hand-side
of the following equation
\begin{equation} \label{BMT11}
\frac{ds^{\mu}}{d\tau}=\Lambda^{\mu}\left[F_{\alpha \beta}, u_{\alpha},\frac{u_{\beta}}{d\tau}, 
s_{\gamma},
...\right]
\end{equation}
where $F_{\alpha \beta}$ is the electromagnetic field-strength tensor.
In deriving $\Lambda^{\mu}$ one makes some assumptions. 
The first one refers to the external fields. 
They should be weak in order to avoid pair-production which is a topic reserved for quantum field
theory. This assumption also tells us that we can restrict ourselves to an expression linear in the
field-strength tensor. The second assumption in connection with the external fields is to assume
the latter changes slowly in time and space. This helps us in as far as we can neglect 
derivatives of
the fields. Finally the third assumption is motivated by (\ref{Heisenbergclassical}): $\Lambda^{\mu}$
should be homogeneous in fields and homogeneous and linear in the spin $s^{\mu}$.

It makes sense to deal first with a general force that is not necessarily of electromagnetic nature
which accelerates the particle. This implies that no field strength tensor should enter our expression 
and therefore,
in agreement with the assumption we made, our candidates for $\Lambda^{\alpha}$ are only two
\begin{equation} \label{1stset}
\Lambda_1^{\alpha}=\{s_{\beta} \frac{du^{\beta}}{d\tau}u^{\alpha}, \epsilon^{\alpha \beta \gamma \lambda}s_{\beta}
\frac{du_{\gamma}}{d\tau} u_{\lambda}\}
\end{equation}
Any other combination either does not satisfy our simple requirements on $\Lambda^{\alpha}$ or is simply 
zero like expression proportional to $\frac{du^{\beta}}{d\tau} u_{\beta}$ 
(this is zero since $u_{\beta}u^{\beta}
=1$). Hence in this general case our relativistic ansatz is simply
\begin{equation} \label{ansatz1}
\frac{ds^{\alpha}}{d\tau}=a\left(s_{\beta}\frac{du^{\beta}}{d\tau}\right) u^{\alpha} 
+b\left (\epsilon^{\alpha \beta \gamma \lambda} s_{\beta}\frac{du_{\gamma}}{d\tau}u_{\lambda}\right)
\end{equation}
Note that in (\ref{ansatz1}) all quantities are to be taken in one and the same frame.
On the other hand the effect of accelerated frame is well known. 
It is called Thomas precession which explicitly evaluated gives \cite{bmt1.2}
\begin{eqnarray} \label{Thomas}
\left(\frac{d \mbox{\boldmath $\xi$}}{dt}\right)_{\rm space}&=&
\left(\frac{d \mbox{\boldmath $\xi$}}{dt}\right)_{\rm rest frame} + \,\,\,
\mbox{\boldmath $\omega$}_T \times 
\mbox{\boldmath $\xi$} \nonumber \\
\mbox{\boldmath $\omega$}_T&=& \frac{\gamma^2}{1 +\gamma}\frac{d {\bf v}}{dt} \times {\bf v}
\end{eqnarray}
with rest frame term given in (\ref{Heisenbergclassical}) and $\gamma=(1-v^2)^{-1/2}$. 
Equation (\ref{Thomas}) is a special version of the fact that $(dG)_{\rm space}=(dG)_{\rm body}
+(dG)_{\rm rotation}$ known from classical mechanics.
Note that this result is universal for any acceleration.
To establish a connection between equations (\ref{ansatz1}) and (\ref{Thomas}) it suffices to use the
Lorentz transformation between $\mbox{\boldmath $\xi$}$ defined in the rest frame and ${\bf s}$ defined
in the same frame where we see the particle moving with velocity ${\bf v}$. We have
\begin{equation} \label{Lorentz}
\mbox{\boldmath $\xi$}={\bf s} -\frac{\gamma}{1 + \gamma}({\bf v}\cdot{\bf s}){\bf v}
\end{equation}
Taking a derivative and using (\ref{ansatz1}) we conclude by comparison with (\ref{Thomas}) that
$a=-1$ and $b=0$.
Admitting in the next step the possibility of acceleration due to electromagnetic force increases
the number of possibilities to be used for $\Lambda^{\alpha}$. Indeed, we get four additional candidates
\begin{equation} \label{2ndset}
\Lambda_2^{\alpha}=\{F^{\alpha \beta} s_{\beta}, s_{\lambda}F^{\lambda \gamma}u_{\gamma}u^{\alpha},
\tilde{F}^{\alpha \beta} s_{\beta}, s_{\lambda}\tilde{F}^{\lambda \gamma}u_{\gamma}u^{\alpha}\}
\end{equation}
with
\begin{equation}
\tilde{F}^{\mu \nu}=\frac{1}{2}\epsilon^{\mu \nu \alpha \beta} F_{\alpha \beta}
\end{equation}
known as the dual electromagnetic field strength tensor which can be obtained from the latter by the 
replacements
${\bf E} \to {\bf B}$ and $ {\bf B} \to -{\bf E}$.
For this case we assume also the validity of Lorentz equation of motion 
\begin{equation} \label{Lorantzeq}
\frac{du^{\alpha}}{d\tau}=\frac{e}{m} F^{\alpha \beta} u_{\beta}
\end{equation} 
Since all non-electromagnetic effects are included in the Thomas precession discussed
above, terms like $s_{\lambda}F^{\lambda \mu} u_{\mu} \frac{du^{\alpha}}{d\tau}$  
and similar
terms with the field strength tensor replaced by its dual are consequently zero for neutral
particles. In case the particle is charged we are
entitled to neglect these terms as they are quadratic in fields. Our most general
ansatz now reads 
\begin{eqnarray} \label{ansatz2}
\frac{ds^{\mu}}{d\tau} &=& AF^{\mu \nu}s_{\nu} + Bu^{\mu}F^{\nu \lambda}u_{\nu} s_{\lambda}
+\tilde{A}\tilde{F}^{\mu \nu}s_{\nu} + \tilde{B}u^{\mu}\tilde{F}^{\nu \lambda}u_{\nu} s_{\lambda}
\nonumber \\
&-& \left(s_{\beta} \frac{du^{\beta}}{d\tau}\right)u^{\mu}
\end{eqnarray}
Note that (\ref{ansatz2}) satisfies automatically the condition
$(ds^{\mu}/d\tau)s_{\mu}=0$ i.e. the conservation of $s^{\mu}s_{\mu}$ as it should be since this 
is already inherent in the non-relativistic equation (\ref{Heisenbergclassical}). This then does 
not give us any new information about the coefficients $A,B,\tilde{A},\tilde{B}$. However,
$(d(s^{\mu}u_{\mu})/d\tau)=0$ tells us that
\begin{equation} \label{firstresult}
A+B=0,\,\,\,\, \tilde{A}+\tilde{B}=0
\end{equation}
The second source of information is the non-relativistic limit of (\ref{ansatz2}) which is
\begin{equation} \label{NR}
\left (\frac{d \mbox{\boldmath $\xi$}}{dt}\right)_{\rm rest frame}
=A \left(\mbox{\boldmath $\xi$} 
\times {\bf B}'\right)
-\tilde{A}\left( \mbox{\boldmath $\xi$}\times {\bf E}'\right)
\end{equation}
Comparing this with (\ref{Heisenbergclassical}) and taking into account
(\ref{firstresult})
we arrive at
\begin{equation} \label{secondresult}
A=-B=2\mu,\,\,\, \tilde{A}=-\tilde{B}=-2d
\end{equation}
which fixes all unknowns in our ansatz. We can now give three different versions of the
generalized BMT equation which as compared to the original version \cite{BMT} 
includes also the electric dipole moment 
$d$. The first version
\begin{eqnarray} \label{BMT1}
\frac{ds^{\mu}}{d\tau} &=& 2\mu\{F^{\mu \nu}s_{\nu} 
- u^{\mu}F^{\nu \lambda}u_{\nu} s_{\lambda}\}
-2d\{\tilde{F}^{\mu \nu}s_{\nu} 
- u^{\mu}\tilde{F}^{\nu \lambda}u_{\nu} s_{\lambda}\}
\nonumber \\
&-& \left(s_{\beta} \frac{du^{\beta}}{d\tau}\right)u^{\mu}
\end{eqnarray}
is the most general one as it is valid for a combination of electromagnetic and non-electromagnetic
forces driving the particle (the non-electromagnetic are contained in the last (Thomas) term).
It is valid for charged as well as neutral particles which at least have non-zero moments 
$\mu$ and $d$.
For charged particles and assuming that the driving force is of electromagnetic
nature only we use the Lorentz force in equation (\ref{Lorentz}) and define
\begin{equation} \label{monents}
\mu=g\frac{e}{4m}, \,\,\, d=-g'\frac{e}{4m}
\end{equation}
Note that we do not use $g=2$ from non-relativistic quantum mechanics as we know already that
$a=g/2-1 \neq 0$. The BMT equation now reads
\begin{equation} \label{BMT2}
\frac{ds^{\mu}}{d\tau} = \frac{e}{m}\left\{\left(\frac{g}{2}\right)F^{\mu \nu}s_{\nu} 
+a u^{\mu}F^{\lambda \nu}u_{\nu} s_{\lambda}
+\left(\frac{g'}{2}\right)\left[\tilde{F}^{\mu \nu}s_{\nu} 
+u^{\mu}\tilde{F}^{\lambda \nu}u_{\nu} s_{\lambda}\right]\right\}
\end{equation}
Finally since the spin is defined in the rest frame of the particle it 
makes sense to use $\mbox{\boldmath $\xi$}$, but to keep ${\bf E}$ and
${\bf B}$ defined in the lab frame. This way one gets the third version
\begin{equation} \label{BMT1a}
\frac{d\mbox{\boldmath $\xi$}}{dt}=\frac{e}{m}
\mbox{\boldmath $\xi$} \times \left(\mbox{\boldmath $\Omega_1$}
+\mbox{\boldmath $\Omega_2$}\right)
\end{equation}
with
\begin{eqnarray} \label{BMT1b}
\mbox{\boldmath $\Omega_1$} &=&
\left( a +\frac{1}{\gamma}\right){\bf B} -a \frac{\gamma}{1+\gamma}
{\bf v}\left({\bf v} \cdot {\bf B}\right) -\left(a +\frac{1}{1+\gamma}\right)
{\bf v} \times {\bf E} \nonumber\\
\mbox{\boldmath $\Omega_2$} &=& 
-\frac{g'}{2}\left[{\bf E} - \frac{\gamma}{1 +\gamma}{\bf v}\left(
{\bf v} \cdot {\bf E}\right) + {\bf v} \times {\bf B}\right]
\end{eqnarray}
Using the last form the spin precession can be investigated in different
field configurations. It is not our objective here to perform such
calculations. Rather we note that essentially the spin precession is closely 
connected to the electromagnetic current through the moments $\mu$ and $d$.
Here the inclusion of the electric dipole moment $d$ is new as compared
to the standard BMT equation \cite{BMT}.  Such a contribution is certainly small
as it violates time reversal symmetry, but still worth a closer 
examination be it only for pedagogical reasons.
Also worthwhile mentioning is the fact that the anomalous magnetic moment
$a$ can be measured using the BMT equation \cite{becher,ll2}.

Certainly, one could also include the precession of the spin due to the
anapole moment starting from a non-relativistic expression $\mbox{\boldmath $\xi$} \times {\bf j}$
where ${\bf j}$ is the current density understood as the source of the electromagnetic fields. i.e.
$\partial_{\mu}F^{\mu \nu}=j^{\nu}$. Evidently such contribution to the spin precession
is possible only in matter. Also obvious is the need to work now with  
derivatives of the fields. Instead of the electromagnetic fields we could, however, work
directly with $j^{\mu}$ to collect all candidates of the corresponding part
of the BMT equation in analogy to what we have done for the magnetic and
electric dipole moment.

\section{Time reversal violating synchrotron radiation}
One of the nice applications of (\ref{BMT1a}) is spin-synchrotron 
radiation in which photons are emitted in the course of a spin transition
from an initial state $i$ to a final one $f$ \cite{jackson}. 
Given the ubiquitous importance of synchrotron radiation in physics and 
astrophysics we consider this as a nice instructive example.
To be able to calculate the 
usual observables of such a radiation we need an effective 
spin-field interaction Hamiltonian. Obviously such an Hamiltonian will 
generalize the non-relativistic results in equations (\ref{int1}) and (\ref{int2}).
The relation between the BMT equation and this Hamiltonian is the same as 
between (\ref{Heisenberg}) and (\ref{Heisenbergclassical}). 
We easily see that we can mathematically exchange in them
the expectation value for the spin operator. Hence using the same technique here
gives us
\begin{equation} \label{hamiltonianrel}
H_{\rm int}^{\rm (eff)}=-\frac{e}{m}\hat{{\bf s}} \cdot \left(
\mbox{\boldmath $\Omega_1$} + \mbox{\boldmath $\Omega_2$}\right)
\end{equation}
Indeed, using the Heisenberg equation with the above Hamiltonian would give 
us the
BMT equation with $\mbox{\boldmath $\xi$}$ replaced by $\hat{{\bf s}}$.
In the spirit of non-relativistic quantum mechanics (not necessarily relativistic
quantum field theory) this Hamiltonian is the relativistic 
generalization of (\ref{hamiltonianrel}) \cite{jackson}. This is part of the reason why it is worth mentioning it.
The standard Hamiltonian for non-spin part of the synchrotron radiation
is
\begin{equation} \label{standardH} 
H_{\rm int} =-e {\bf A}\cdot {\bf v}
\end{equation}
where vector potential of the photon field is
\begin{equation} \label{A}
{\bf A}= \mbox{\boldmath $\epsilon$}\kappa \exp\{i{\bf k}\cdot{\bf r}
-i\omega t \} +{\rm c.c.}
\end{equation}
where $\kappa$ is often chosen to be $(2\pi/\omega)^{1/2}$.
The fields entering (\ref{hamiltonianrel}) can be easily calculated using (\ref{A}).
Since we are discussing photon emission we need to consider only the
complex conjugate part of (\ref{A}). 
The final result is
\begin{eqnarray} \label{final}
[H_{\rm int}^{\rm (eff)}]_{\rm emission}&=&i\sqrt{2\pi \omega}\,\,\,
\hat{{\bf s}} \cdot \left({\bf V}_1 +{\bf V}_2\right)e^{i\omega t-i{\bf k}\cdot {\bf r}}
\nonumber \\
{\bf V}_1 &=&
\left(a +\frac{1}{\gamma}\right) \hat{{\bf k}} \times \mbox{\boldmath $\epsilon$}^*
- a\frac{\gamma}{1 +\gamma}{\bf v}{\bf v}\cdot (\hat{{\bf k}} \times \mbox{\boldmath $\epsilon$}^*)
-\left(a +\frac{1}{1+\gamma}\right){\bf v}\times \mbox{\boldmath $\epsilon$}^*
\nonumber \\
{\bf V}_2 &=& -\frac{g'}{2}\left[ \mbox{\boldmath $\epsilon$}^*
-\frac{\gamma}{1+\gamma}{\bf v}({\bf v}\cdot \mbox{\boldmath $\epsilon$}^*)
+ {\bf v} \times (\hat{{\bf k}} \times \mbox{\boldmath $\epsilon$}^*)\right]
\end{eqnarray}
Note that in contrast to (\ref{standardH})
the Hamiltonian (\ref{hamiltonianrel}) and (\ref{final}) obviously contain the spin 
$\hat{\bf s}=\hbar\mbox{\boldmath $\sigma$}/2$.
Here we have re-introduced $\hbar$ explicitly (in the rest of the paper we set
$\hbar=1$) to make clear that the effects calculated with the help of (\ref{hamiltonianrel}) are
'true' quantum mechanical effects (in contrast to standard, classical synchrotron radiation 
effects which
are zeroth order in $\hbar$). An example of such an effect is e.g. the polarization  
of electrons and/or positrons  
in storage rings which is explained by using the 
$\mbox{\boldmath $\Omega$}_1$ term in (\ref{hamiltonianrel}). The 
calculation of the effect follows the methods of quantum field theory, however using
(\ref{hamiltonianrel}) seems to be more instructive as it is done 
at the level of quantum theory and what is more, it contains 
an arbitrary $g$ factor which helps to explain certain issues of the result. 

The part proportional to $\mbox{\boldmath $\Omega$}_2$ in (\ref{hamiltonianrel})
violates time reversal symmetry.
This is seen
that by adding to our previous $T$-transformations (\ref{Ttrans}) the obvious rule
\begin{equation} \label{TT}
{\bf v} \xrightarrow{\ T} -{\bf v}\end{equation}
As discussed in connection with the electric dipole
moment itself we know that phenomena which violate time reversal symmetry are extremely rare.
We therefore 
expect that the measurable effects in connection with this part of the Hamiltonian
are also very small. However, there is also a rewarding aspect of our Hamiltonian.
Again time-reversal violating effects are introduced using the machinery
of quantum field theory \cite{ll3}. Here we derived one example by very simple means.
In principle, it must be possible to construct an observable in the form of an asymmetry
to extract the time reversal violating part i.e. these effects should be
proportional only to $d$.  
This can serve as an introduction to important
tools used all over physical sciences.
The explicit calculation of the rate for synchrotron radiation with spin transition
can be carried through along the same lines as explained 
in reference \cite{jackson}, where only the first part
of the Hamiltonian (\ref{final}) was included.
\section{Discussion}
It is a long way to begin with the Dirac equation and the relativistic current for a spin-1/2
fermion and arrive at the time-reversal violating synchrotron radiation. 
This demonstrates the rich phenomenology associated with the current
and the form factors contained in it. 
From the discrepancy between different measurements of $F_{i=1,2}$ for
the nucleons and from the absence of
information on $F_{i=3,4}$, we can see that the subject is still 
an active field for investigations. Seen from this point of view, one could say that 
the 
form factors of the nucleons are still not accurately known, a rather surprising and disturbing
conclusion, as the form factors are an important source of information on the nucleon
structure \cite{nucleon}.
For this reason we were motivated to present in this article a comprehensive summary of the subject.

The new term (\ref{pauliterm}) which enters the non-relativistic Pauli current mentioned in the introduction
is not the end of the story. Indeed, we have shown that three other terms are also possible in the
relativistic formulation.

All the physics presented in the paper can be understood by students
who have mastered the Dirac equation. A large part of the phenomenology is still
unexplored and as such good material for students to do some
research and projects of their own. 
This is certainly the case when we look for solutions of the BMT equation in various
field configurations. The time reversal violating part also offers several
possibilities in this direction. A BMT equation including the anapole moment remains 
a subject open to investigation.

A good deal of physics and its different methods of calculation and reasoning can be 
introduced by deriving the current and its consequences, as done in the present article. 
Some example are the
discrete symmetries and their violation. Another example is the
derivation of the BMT equation which proceeds via elimination of candidates.
Features important for fundamental physics and astrophysics,  
such as the difference between Majorana and Dirac fermions,  
have been touched upon.

Finally, a quantum mechanical current has, of course, 
many different applications, sometimes surprising ones \cite{me2}.

\appendix
\addcontentsline{toc}{section}{Appendices}
\section{Appendix: The Dirac Algebra and Gordon Identities} 
The following appendix is a little formal, but since the formulae which we will gather in it
will help us to pin down the general electromagnetic current it is worth to go
through the algebra. The reader who is familiar with Dirac algebra and/or wishes to follow 
more directly the flow
how the general electromagnetic current is obtained can skip the appendix and proceed 
with the main text without losing the track and the physics of the
electromagnetic form factors.

We assume
that the reader is familiar with the Dirac equation (see \cite{bj1, itzykson} for details):
\begin{equation}  \label{dirac}
(i\ing\partial-m)\Psi (x) =0,
\end{equation}
as well as, with the customary notation of `slashing' a four-vector $a_{\mu}$ i.e. 
$\ing\hspace{-3pt} a=a_{\mu}\gamma^{\mu}$(hence $\ing\hspace{-3pt} 
\partial=\partial_{\mu}\gamma^{\mu}$) 
and the `bar-notation' i.e. $\bar{\Psi}=\Psi^{\dagger} \gamma^0$.
We will keep the following summary of the Dirac-equation and its properties short
since it is described in many text books like \cite{bj1} and \cite{itzykson}. 
The positive energy plane wave solution
of (\ref{dirac}) is 
\begin{equation}\label{solve+e}
\Psi(x)=u({\bf p})e^{-ipx}
\end{equation}
which amounts to solving the  Dirac equation in momentum space
\begin{equation}  \label{+eplanewave}
(\ing p-m)u ({\bf p}) =0
\end{equation}
We now summarize some important properties of the $\gamma_{\mu}$ algebra:
\begin{equation}
\label{propertiesofgamma}
\begin{gathered} 
\{ \gamma^{\mu},\gamma^{\nu}\}=2g^{\mu\nu}\quad  ;\quad \left\{ \gamma^5 ,\gamma^{\nu} \right\}=0\\
\gamma_5  \equiv  -\frac{i}{4!}\epsilon_{\mu\nu\alpha\beta}
\gamma^{\mu}\gamma^{\nu}\gamma^{\alpha}\gamma^{\beta}\\
\sigma^{\mu\nu}\equiv \frac{i}{2}[\gamma^{\mu} ,\gamma^{\nu}]\quad ;\quad \gamma^{\mu}
\gamma^{\nu}=g^{\mu\nu}-i\sigma^{\mu\nu}\\
\gamma_5 \sigma^{\mu\nu}=\frac{i}{2}\epsilon^{\mu\nu\alpha\beta} \sigma_{\alpha\beta}\\
[\gamma_5 ,\sigma^{\mu\nu}]=0
\end{gathered}
\end{equation}
and obviously we obtain $\gamma_5 \sigma^{\mu \nu}q_{\nu}=\frac{i}{2}\epsilon^{\mu \nu \alpha \beta}q_{\nu}$ 
which enters the set {\ref{o3}).
It is well known \cite{itzykson} that the set $\mathcal{S}$, where
\begin{equation}\label{gammabasis}
\mathcal{S}=\{\1,\gamma_5,\gamma_{\mu},\gamma_5\gamma_{\mu},\sigma_{\mu\nu}\}
\end{equation}
forms the basis of the 16 linearly independent, with the exception of the unit matrix, traceless $4\times4$ matrices. 
For any member $\Gamma \in \mathcal{S}$ we have

\begin{eqnarray}\label{daggergamma}
\Gamma^\dagger&=&\eta_0[\Gamma]\gamma^0\Gamma\gamma^0\notag\\
\Gamma^{\rm T}&=&\eta_T[\Gamma] C \Gamma C^{-1}
\end{eqnarray}
where $C$ is the charge conjugation matrix and $\eta_0[\Gamma]$ as well as $\eta_T[\Gamma]$ are
pure signs depending on $\Gamma\in\mathcal{S}$. 
In the same order as in (\ref{gammabasis}) they are given by
$\eta_0[\Gamma]=(+,-,+,+,+)$ and $\eta_T[\Gamma]=(+,+,-,+,-)$.
The reader who wishes  to refresh 
her or his  memory on Dirac algebra should consult the appendix of \cite{itzykson}. 
Finally, we will make use of 
Gordon-like identities easily derivable using (\ref{solve+e}) and (\ref{propertiesofgamma}):
\begin{eqnarray}\label{gordonlike1}
\bar{u}({\bf p}_1)\gamma^{\mu} u({\bf p}_2)&=&\frac{1}{2m}\bar{u}({\bf p}_1)[l^\mu 
+ i\sigma^{\mu\nu}q_{\nu}]u({\bf p}_2)\notag\\
\bar{u}({\bf p}_1)\gamma^{\mu}\gamma_5 u({\bf p}_2)&=&\frac{1}{2m}\bar{u}({\bf p}_1)
[\gamma_5 q^\mu + i\gamma_5 \sigma^{\mu\nu}l_{\nu}]u({\bf p}_2)\notag\\
\bar{u}({\bf p}_1)i\sigma^{\mu\nu}l_{\nu} u({\bf p}_2)&=&-\bar{u}({\bf p}_1)q^{\nu}u({\bf p}_2)\notag\\
\bar{u}({\bf p}_1)i\sigma^{\mu\nu}q_{\nu} u({\bf p}_2)&=&\bar{u}({\bf p}_1)[2m \gamma^{\mu}l^{\mu}]
u({\bf p}_2)
\end{eqnarray}
The second set of similar identities involves the Levi-Cevita tensor (recall the connection
between the Levi-Cevita tensor and the $\gamma_5 \sigma_{\mu \nu}$ product 
in (\ref{propertiesofgamma}))
\begin{eqnarray} \label{gordonlike2}
\bar{u}({\bf p}_1)i\sigma^{\mu\nu}\gamma_5 q_{\nu} u({\bf p}_2)
&=&-\bar{u}({\bf p}_1)l^{\mu}\gamma_5 u({\bf p}_2)\notag \\
\bar{u}({\bf p}_1)[\epsilon^{\alpha\mu\nu\beta}\gamma_5\gamma_{\beta}q_{\mu}l_{\nu}] 
u({\bf p}_2)&=&\bar{u}({\bf p}_1)\{-i[q^{\alpha}\ingl - l^{\alpha} \ing q]+i(q^2-4m^2)\gamma^{\alpha} +\notag\\
& &2im(l^{\alpha}+q^{\alpha}) \}u({\bf p}_2)\\
\bar{u}({\bf p}_1)[\epsilon^{\alpha\mu\nu\beta}\gamma_{\beta}q_{\mu}l_{\nu}] 
u({\bf p}_2)&=&\bar{u}({\bf p}_1)\{i[q^{\alpha}\ingl - l^{\alpha}\ing q]\gamma_5+iq^2
\gamma_5\gamma^{\alpha}-\notag\\
& &2im(l^{\alpha}+q^{\alpha})\gamma_5 \}u({\bf p}_2)\notag\\
\bar{u}({\bf p}_1)[\epsilon^{\mu\nu\alpha\beta} q_\alpha l_\beta\gamma_{\nu}\gamma_5] 
u({\bf p}_2)&=&\frac{i}{2m}\bar{u}({\bf p}_1)[\epsilon^{\mu\nu\alpha\beta} 
q_\alpha l_\beta\sigma_{\nu\rho}q^\rho]u({\bf p}_2)\notag\\
\bar{u}({\bf p}_1)[\epsilon^{\mu\nu\alpha\beta} q_\alpha l_\beta\sigma_{\nu\rho}l^\rho]
u({\bf p}_2)&=&0\notag
\end{eqnarray}
We have defined 
\begin{eqnarray}
l^{\alpha}& \equiv & p_1^{\alpha}+ p_2^{\alpha}\notag \\
q^{\alpha}& \equiv & p_1^{\alpha}- p_2^{\alpha}.
\end{eqnarray}

We offer here a proof of the second  identity in (\ref{gordonlike2}) and suggest the rest as an exercise 
to the reader. The first step in proving this identity is to use the basis given in 
(\ref{gammabasis}). In 
particular it means that any product of two or more matrices from $\mathcal{S}$ can be 
expanded into the basis of $\mathcal{S}$. Given $\Gamma_i , \Gamma_j \in \mathcal{S}
\textrm{ for } i \not= j$ we can convince ourselves that ${\rm Tr}[\Gamma_i \Gamma_j]=0$. Hence 
the coefficients of the expansion are easily calculable. For instance:

\begin{equation}
\gamma^{\alpha} \sigma^{\mu\nu}=a^{\mu\nu\alpha} \1 
+b^{\mu\nu\alpha}\gamma_5 +c^{\mu\nu\alpha\beta}\gamma_{\beta} 
+d^{\mu\nu\alpha\beta}\gamma_5\gamma_{\beta} +e^{\mu\nu\alpha\rho\delta}\sigma_{\rho\delta}
\end{equation}
We have
\begin{equation*}
{\rm Tr}[\gamma^{\alpha} \sigma^{\mu\nu}]=a^{\mu\nu\alpha} {\rm Tr}[\1],
\end{equation*}
and
\begin{equation*}
{\rm Tr}[\gamma^{\alpha} \sigma^{\mu\nu}]=0, \textrm{ hence }a^{\mu\nu\alpha}=0.
\end{equation*}
Similarly, from
\begin{equation*}
{\rm Tr}[\gamma^{\alpha} \sigma^{\mu\nu}\gamma_5]
={\rm Tr}[\gamma^{\alpha} \sigma^{\mu\nu}\sigma_{\rho\delta}]=0
\end{equation*}
it follows that
\begin{equation*}
b^{\mu\nu\alpha}=e^{\mu\nu\alpha\rho\delta}=0.
\end{equation*}
The non-zero coefficient can be computed by noticing that
\begin{equation*}
\begin{gathered}
{\rm Tr}[\gamma^{\alpha} \sigma^{\mu\nu}\gamma^{\beta}]=4c^{\mu\nu\alpha\beta} \quad \textrm{and} \\
{\rm Tr}[\gamma_5 \gamma^{\alpha} \sigma^{\mu\nu}\gamma^{\beta}]=4d^{\mu\nu\alpha\beta}
\end{gathered}
\end{equation*} 
Calculating these traces explicitly we obtain 
\begin{equation}\label{gammasigma}
\gamma^{\alpha} \sigma^{\mu\nu}=i[g^{\alpha\mu} g^{\beta\nu}-g^{\alpha\nu} g^{\beta\mu}]\gamma_{\beta} 
+ \epsilon^{\alpha\mu\nu\beta} \gamma_5 \gamma_{\beta}
\end{equation}
The easy step is to contract (\ref{gammasigma}) with $q_{\mu} l_{\nu}$ and sandwich 
this between $\bar{u}({\bf p}_1)$ and $u({\bf p}_2)$. This leads to 
\begin{equation}
\bar{u}({\bf p}_1)[\epsilon^{\alpha\mu\nu\beta}\gamma_5\gamma_{\beta}q_{\mu}l_{\nu}] 
u({\bf p}_2)=-i\bar{u}({\bf p}_1)[q^{\alpha}\ingl - l^{\alpha} \ing q]u({\bf p}_2)
+\bar{u} ({\bf p}_1)[\gamma^{\alpha}\sigma^{\mu\nu}q_{\mu}l_{\nu})]u({\bf p}_2)\notag\\
\end{equation}
The last step is to note that 
\begin{equation*}
-2i\sigma^{\mu\nu}q_{\mu}l_{\nu}= \ing q \ingl - \ingl \ing q = -2q\cdot l +2\ing q \ingl 
= 2\ing q \ingl\textrm{ as } q \cdot l=0
\end{equation*} 
We can expand $\ing q \ingl$ i.e.
\begin{equation*}
\ing q \ingl =- \ing p_2 \ing p_1  +\ing p_1 \ing p_2 = 2(-p_1\cdot p_2 +\ing p_1 \ing p_2) 
= 2\left( \frac{q^2 -2m^2}{2}+\ing p_1  \ing p_2\right)
\end{equation*} 
and use (\ref{+eplanewave}) to arrive at the second identity in (\ref{gordonlike2})


\begin{thebibliography}{99}
\bibitem{note1}
More precisely, in an electromagnetic theory the current conservation is equivalent to demanding
gauge invariance which in turn implies that the mass of the corresponding gauge boson i.e. the photon
be zero. Since the gauge invariance in the weak sector is (spontaneously) broken, the corresponding 
weak gauge bosons become massive and as a result we do not have a conserved weak current.
\end{thebibliography}
\end{document}